\documentclass[prl,preprint,aps,longbibliography,groupedaddress]{revtex4-1}

\usepackage{bbm}
\usepackage{graphicx}
\usepackage{subfigure}
\usepackage{amsmath}
\usepackage{booktabs}

\begin{document}

\title{Observation of diamagnetic strange-metal phase in sulfur-copper codoped lead apatite}

\author{Hongyang Wang$^{1}$\footnote{\url{wanghy@ipe.ac.cn}}, Hao Wu$^{2}$, Ning Chen$^{3}$, Xianfeng Qiao$^{4}$, Ling Wang$^{5}$, Zhixing Wu$^{6}$, Zhihui Geng$^{7}$, Weiwei Xue$^{8}$, Shufeng Ye$^{1}$, and Yao Yao$^{4,9}$\footnote{\url{yaoyao2016@scut.edu.cn}}}

\address{$^1$ Center of Materials Science and Optoelectronics Engineering, Institute of Process Engineering, Chinese Academy of Sciences, Beijing 100049, China\\
$^2$ School of Materials Science and Engineering, Huazhong University of Science and Technology, Wuhan 430074, China\\
$^3$ School of Materials Science and Engineering, University of Science and Technology Beijing, Beijing 100083, China\\
$^4$ State Key Laboratory of Luminescent Materials and Devices, South China University of Technology, Guangzhou 510640, China\\
$^5$ Beijing Key Laboratory of Ionic Liquids Clean Process, Institute of Process Engineering, Chinese Academy of Sciences, Beijing 100190, China\\
$^6$ Fujian Provincial Key Laboratory of Analysis and Detection Technology for Food safety, College of Chemistry, Fuzhou University, Fuzhou 350108, China\\
$^7$ School of Engineering, Course of Applied Science, Tokai University, Hiratsuka 2591292, Japan\\
$^8$ CAS Key Laboratory of Mechanical Behavior and Design of Materials, Department of Modern Mechanics, University of Science and Technology of China, Hefei 230027, China\\
$^9$ Department of Physics, South China University of Technology, Guangzhou 510640, China}

\date{\today}

\begin{abstract}
By codoping sulfur and copper into lead apatite, the crystal grains are directionally stacked and the room-temperature resistivity is reduced from insulating to $2\times10^{-5}~\Omega\cdot$m. The resistance-temperature curve exhibits a nearly linear relationship at low temperature suggesting the presence of strange-metal phase, and a second-order phase transition is then observed at around 230~K during cooling the samples. A possible Meissner effect is present in dc magnetic measurements. Further hydrothermal lead-free synthesis results in smaller resistance and stronger diamagnetism, demonstrating the essential component might be sulfur-substituted copper apatite and the alkalis matter as well. A clear pathway towards superconductivity in this material is subsequently benchmarked.
\end{abstract}

\maketitle

Charge carriers in normal metal are described as quasiparticles of elementary excitations in Fermi liquid theory, from which the resistance--temperature (RT) relationship basically follows quadratic form. A strange metal however possesses a linear-in-temperature resistivity, long-termly challenging the traditional understanding of charge transport \cite{2020natStrange, 2021natsm, 2023scistrange, 2023npstrange,2021ncstrange}. In particular, as the normal phase of cuprate or iron-based superconductors has often been strange metal, it is then intuitively related to the strongly correlated electrons \cite{2023npcuprate, 2019npcuprate, 2023natiron}. The strange metal also appears in the Bechgaard salt, a quasi-one-dimensional (1D) organic metal, implying it prefers to low dimensions \cite{2022natBechgaard,2009prbBechgaard}. As a quantum critical phase, the strange metal is regarded to be spatially ordered so with good conductivity, but temporally disordered so without global phase coherence \cite{prlphase_coherence}. The anisotropy in low dimension gives rise to local critical fluctuations which results in dissociation of vortex pairs in superconducting phase \cite{prbvortex_pairs}. An individual vortex could thus be detectable in the quasi-ordered strange-metal phase.

Materials with multiple elements, such as cuprate superconductor, make diversity in physics. It is always difficult to deal with multiple elements, since each of them is individually activated at distinct temperature. Copper-substituted lead apatite, also named as LK-99 \cite{Lee1,Lee2}, was claimed to be room-temperature superconductor, but due to the complicated components and structures, the reproduction is still controversial \cite{Wang2023,Guo2023,2023habamahoro,liu2023longcoherence,2023lowfield,wang2024possible}. Considering copper is a sulfophilic element, the participation of sulfur in the synthesis benefits the substitution of copper, which does originally not favor the ionic crystal structure of lead apatite. As reported in our previous work \cite{wang2024possible}, this sulfur-copper codoped lead apatite (SCCLA) manifests a weak Meissner effect at near room temperature. In order to further enhance the effect, we have to either finely optimize the reaction procedure as sulfur could not be held in the bulk at overhigh temperature that enables other elements to react, or alternatively design and synthesize some lead-free frameworks. In this work, we then adjust the synthetic procedure of SCCLA and find the signal magnitude is largely increased.

\begin{figure*}[t]
    \centering
    \includegraphics[width=0.9\linewidth]{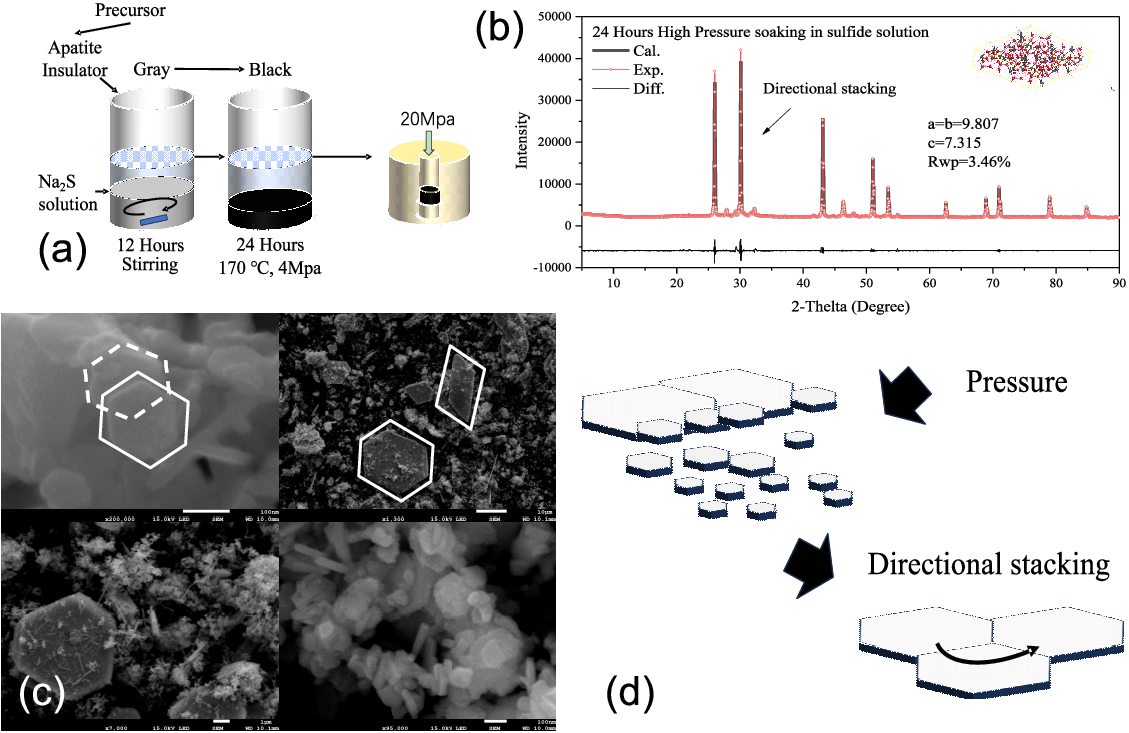}
    \caption{(a) Schematic of sample synthesis procedure of sulfur doping in SCCLA. (b) XRD pattern of the S1 sample after 24 hours soaking in sulfide solution. (c) The morphological characteristics with the hexagonal grains highlighted. (d) Schematic of the directional stacking mechanism under high pressure. }
\end{figure*}

Pure copper-substituted lead apatite was first synthesized through a hydrothermal-calcination method following the procedure outlined in the literature \cite{Lee1,wang2024possible}, with the resulting gray powder being insulating. In order to carry out sulfur doping, as sketched in Fig.~1(a), the powder is ball-milled to nano-scaled particles and then thoroughly dispersed in a saturated solution of sodium sulfide. After stirring for 12 hours, the sample undergoes a 24-hour reaction in a high-pressure reaction vessel. The internal pressure of the reaction vessel is maintained at no less than 4~MPa, and the reaction temperature is set to 170~$^{\circ}$C. After the reaction, the powder is washed, filtered, dried, and then subject to a pressure of 20~MPa to compress it into a solid tablet. The resulting samples labeled as S1 and S2 are parallel with the latter was purified more carefully.

The XRD spectra of S1 sample are refined using Reflex as shown in Fig.~1(b). Pawley Fitting is utilized to accurately determine lattice constants, and experimental and sample-related parameters. During the sulfur doping process, atoms within the apatite structure may have been etched or consumed, so the lattice is largely shrunk leading to a strong lattice distortion and good conductivity. The experimental spectra are consistent with that of the phosphate of P63m structure. By comparing to the standard of lead phosphate, the intensity differences observed at some peak positions could be attributed to the orientation of nano-scaled grains during growth in the sulfide solution. SEM was utilized to observe the morphological characteristics of the sample, as displayed in Fig.~1(c). The powder mainly consists of hexagonal flakes or hexagonal crystals. A small amount of hexagonal prism rod-like particle shows up, pointing out a significant difference compared to the sample before sulfur doping. This observation together with the analysis results of XRD spectra reveal the microstructure of the powder emerging as hexagonal grains, so the directional stacking is evident to function during high-pressure agglomeration, as sketched in Fig.~1(d). The head-to-tail quasi-1D stacking structure is then changed to be side-by-side, so the conducting channel might be broadened.

\begin{figure*}[t]
    \centering
    \includegraphics[width=0.9\linewidth]{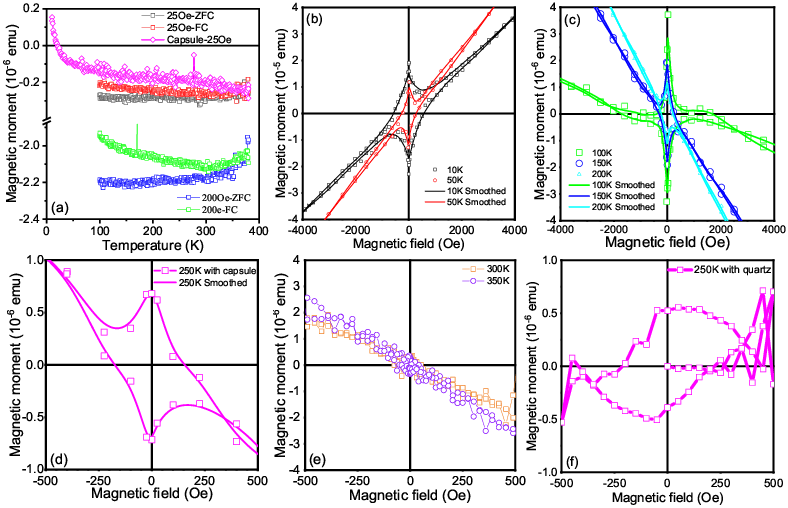}
    \caption{(a) MT curves of ZFC and FC measurements of S1 sample at 25 and 200~Oe, respectively. The signal of an empty capsule is shown for comparisons. (b)-(e) MH hysteresis curves at eight temperatures. At 10 and 50~K, it is paramagnetic at strong field, and at higher temperature it becomes diamagnetic. Below 250~K, a remarkable magnetic hysteresis effect is observed between $-300$ and 300~Oe. At 300 and 350~K, although we have decreased the measurement interval, there is not visible hysteresis indicating the critical temperature is below 300~K. (f) MH hysteresis curve at 250~K is again finely measured with a quartz holder. The slight difference of lineshape may be due to the orientational difference of tablet samples on different holders.}
\end{figure*}

We then employed MPMS-3 SQUID to measure the dc magnetic moments of S1 sample, as shown in Fig.~2. It is clear from the magnetic moment--temperature (MT) curves, considering the background of capsule, the magnetization under magnetic field of 25 and 200~Oe is negative up to near room temperature, indicating its diamagnetism at weak field. The bifurcation between zero-field-cooling (ZFC) and field-cooling (FC) measurements is also found to be above 250~K. The moment--magnetic field (MH) curves at various temperatures from 10 to 350~K are also displayed. One can see that, from $-300$ to 300~Oe, the MH curves exhibit notable hysteresis effect up to 250~K. In particular, the hysteresis at 150~K is pretty obvious, exceeding the highest critical temperature of known superconductors at ambient pressure. This phenomenon has been reported in our previous paper \cite{wang2024possible}, but the quality of the present data is largely increased, further eliminating the possibility of measurement faults. At 300 and 350~K, the hysteresis can not be detected. As stated before, these results strongly suggest there exists a Meissner effect in SCCLA.

\begin{figure*}[t]
    \centering
    \includegraphics[width=0.9\linewidth]{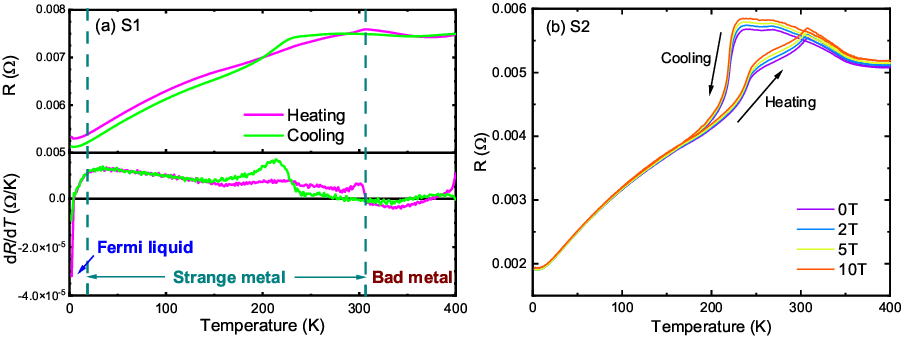}
    \caption{(a) The RT curves and their derivative at current of 8~mA during heating and cooling the S1 sample. Based on the RT relationship, the temperature region can be divided into three: Fermi liquid, strange metal and bad metal. (b) RT curves of S2 sample at various magnetic field.}
\end{figure*}

Measurement results of transport with indium electrode are displayed in Fig.~3, which are mainly conducted on PPMS and have been reconfirmed on other facilities. Although the sample before soaking in sulfur is insulator, after soaking the resistivity at room temperature is largely reduced to around $2\times10^{-5}~\Omega\cdot$m, close to that of natural graphite. It is thus evident that the sulfur-copper codoping plays essential role in the improvement of transport properties of the insulating apatite ionic crystal. Such considerable conductivity is far beyond expectation, that has to be carefully comprehended.

As estimated by the critical magnetic field, if there is a superconducting phase, the critical current should be as small as $\mu$A, which can not be accurately detected in our present facility. The RT curves are therefore measured with constant current of 8~mA to ensure the precision, as illustrated in Fig.~3. It is found that below 20~K the derivative $dR/dT$ exhibits a nearly linear relation suggesting a Fermi liquid behavior as in normal metals. Increasing the temperature, the derivative becomes nearly constant revealing a linear-in-temperature resistivity, which is an essential signature of strange metal. At around 230~K in the cooling curve and 304K in the heating curve, there are remarkable turning points to figure out the transition from strange metal to bad metal phase. The transition is more evident in S2 sample than that in S1, implying it is a phase transition of second order. It has been stated that, if the resistivity in bad metal can be further reduced by one or two orders \cite{2022natBechgaard}, these transition points can then be recognized as the superconducting critical temperature. The thermal hysteresis of the two RT curves, probably stemming from the poor thermal conductivity of the samples, is also consistent with that of the magnetic measurements.

\begin{figure*}[t]
    \centering
    \includegraphics[width=0.99\linewidth]{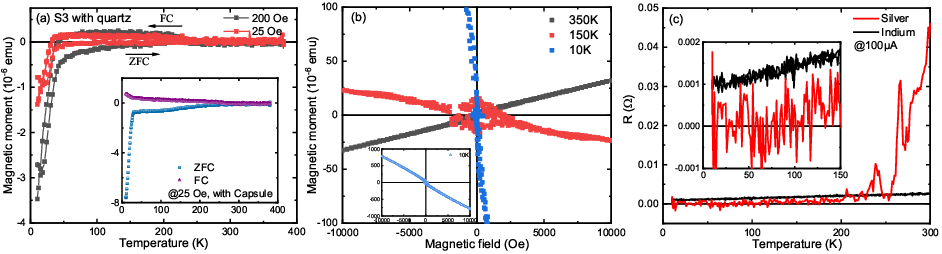}
    \caption{(a) MT curves of S3 sample with quartz holder at 25 and 200~Oe for ZFC and FC measurements. Inset displays the MT curves at 25~Oe with capsule. The bifurcation occurs around 250~K. (b) MH curves at three typical temperatures. The diamagnetism is too strong at 10~K, so it is also displayed with larger scale in the inset. (c) RT curves at 100~$\mu$A with the electrodes being silver and indium, respectively. Inset shows the temperature range within 0 to 150~K. At low temperature, the resistance of silver electrode sample is too small to be detectable due to the limit of instrument.}
\end{figure*}

On the current stage, we are wondering which component dominates the strange features of SCCLA. In particular, the lead seems to be inactive in all the measurements. To this end, we further synthesized a lead-free compound labeled as S3, solely by hydrothermal method with the feed ratio of Cu$^{2+}$ and PO$^{3-}_4$ equal to $10:6$ and the sodium being replaced by potassium. The replacement of alkali metals could further enhance the conductivity but does not introduce other qualitative changes. The magnetic and electric measurement results are displayed in Fig.~4, respectively. It is found that, both MT and MH curves show similar features with that of SCCLA, that is, a possible superconducting transition occurs at around 250~K. This suggests the lead does not matter in the superconductivity, and as the lead-free sample is more fragile than that with lead, the role of lead might merely be to enhance the robustness of structure. Interestingly, below 40~K, the magnetic moment drops dramatically by more than one order. The different FC moment with different holders at low temperature may be due to the orientational difference of sample under magnetic field, as the crystal grains have been directionally stacked. By MH curve at 10~K, it is also observed that the diamagnetism is greatly enhanced, far exceeding the normal diamagnetism in metals, which can only be explained by superconductivity.

The RT curves of the same sample are individually measured with different electrodes, silver and indium. It is quite strange that, with silver the room-temperature resistance is much larger than that with indium, but at low temperature, the resistance with silver becomes much smaller, even undetectable. The transition point with silver electrode is also around 250~K, consistent with that of diamagnetism, excluding the possibility of virtual connection. It is noticed that, after measurement, the color of silver electrode becomes black implying it has been sulfurized. We therefore think that, the silver extracts the sulfur in the 1D ionic channel of apatite, leading to an electron-rich state and thus strengthening the superconductivity. The indium electrode can however not closely contact to the apatite structure, so it solely performs a strange-metal behavior.

Compared to our previous synthetic procedure \cite{wang2024possible}, the doping of sulfur not only facilitates the performance of copper, but also changes the topology of 1D ionic channel of apatite. From head-to-tail to side-by-side, the quasi-1D copper-sulfur lattice in apatite may have got much stronger inter-chain interaction to activate tunneling in between. The alkali metals may also play roles in connecting the crystal grains. This explains the huge improvement of electric conductivity and the emergence of Meissner effect. We have also conducted microwave absorption measurement and did not find visible radical signals, implying the absence of quasiparticles in normal metal. People might suspect the magnetic hysteresis is enabled by some pseudogap states, but since the RT curves at least exhibit a metallic feature and extremely low resistance in lead-free samples, we would be more preferable to think it is because the superconducting critical current is below our measurement limit. During measurements, we have noticed that the SCCLA sample possesses abnormally large electric capacity and poor thermal conductivity \cite{2022cap,2004cap}, which are inconsistent with the high electric conductivity. In particular, the capacity often induces strange sudden change between low- and high-resistance states enabling large fluctuations of RT curves, even if we have carefully excluded other interferences. It is more likely the sample is continuously charging and discharging during electric measurements, hindering the detectability of zero resistance in SCCLA. Considering the performance of lead-free sample, the dielectrics of lead apatite backbone matters.

In summary, we modify the synthetic procedure of SCCLA to codope both sulfur and copper into lead apatite, and the structural characterization reveals a directional stacking mechanism. The magnetic and electric properties of SCCLA have been comprehensively investigated. The hysteresis MH loops can be observed up to 250~K, and the ZFC--FC bifurcation occurs as well. The RT curve manifests that SCCLA possesses a strange-metal phase at large current and a second-order phase transition occurs at around 230~K during cooling. Further synthesis of lead-free sample is then performed which provides us even stronger diamagnetism and smaller resistance at low temperature. We therefore believe that we have made a substantial step towards room-temperature superconductivity.

\section{Acknowledgments}

The authors gratefully acknowledge support from the National Natural Science Foundation of China (Grant Nos.~12374107 and 52304430).

\bibliography{Strange_v4.bbl}

\begin{thebibliography}{22}%
\makeatletter
\providecommand \@ifxundefined [1]{%
 \@ifx{#1\undefined}
}%
\providecommand \@ifnum [1]{%
 \ifnum #1\expandafter \@firstoftwo
 \else \expandafter \@secondoftwo
 \fi
}%
\providecommand \@ifx [1]{%
 \ifx #1\expandafter \@firstoftwo
 \else \expandafter \@secondoftwo
 \fi
}%
\providecommand \natexlab [1]{#1}%
\providecommand \enquote  [1]{``#1''}%
\providecommand \bibnamefont  [1]{#1}%
\providecommand \bibfnamefont [1]{#1}%
\providecommand \citenamefont [1]{#1}%
\providecommand \href@noop [0]{\@secondoftwo}%
\providecommand \href [0]{\begingroup \@sanitize@url \@href}%
\providecommand \@href[1]{\@@startlink{#1}\@@href}%
\providecommand \@@href[1]{\endgroup#1\@@endlink}%
\providecommand \@sanitize@url [0]{\catcode `\\12\catcode `\$12\catcode
  `\&12\catcode `\#12\catcode `\^12\catcode `\_12\catcode `\%12\relax}%
\providecommand \@@startlink[1]{}%
\providecommand \@@endlink[0]{}%
\providecommand \url  [0]{\begingroup\@sanitize@url \@url }%
\providecommand \@url [1]{\endgroup\@href {#1}{\urlprefix }}%
\providecommand \urlprefix  [0]{URL }%
\providecommand \Eprint [0]{\href }%
\providecommand \doibase [0]{http://dx.doi.org/}%
\providecommand \selectlanguage [0]{\@gobble}%
\providecommand \bibinfo  [0]{\@secondoftwo}%
\providecommand \bibfield  [0]{\@secondoftwo}%
\providecommand \translation [1]{[#1]}%
\providecommand \BibitemOpen [0]{}%
\providecommand \bibitemStop [0]{}%
\providecommand \bibitemNoStop [0]{.\EOS\space}%
\providecommand \EOS [0]{\spacefactor3000\relax}%
\providecommand \BibitemShut  [1]{\csname bibitem#1\endcsname}%
\let\auto@bib@innerbib\@empty
%</preamble>
\bibitem [{\citenamefont {Shen}\ \emph {et~al.}(2020)\citenamefont {Shen},
  \citenamefont {Zhang}, \citenamefont {Komijani}, \citenamefont {Nicklas},
  \citenamefont {Borth}, \citenamefont {Wang}, \citenamefont {Chen},
  \citenamefont {Nie}, \citenamefont {Li}, \citenamefont {Lu}, \citenamefont
  {Lee}, \citenamefont {Smidman}, \citenamefont {Steglich}, \citenamefont
  {Coleman},\ and\ \citenamefont {Yuan}}]{2020natStrange}%
  \BibitemOpen
  \bibfield  {author} {\bibinfo {author} {\bibfnamefont {Bin}\ \bibnamefont
  {Shen}}, \bibinfo {author} {\bibfnamefont {Yongjun}\ \bibnamefont {Zhang}},
  \bibinfo {author} {\bibfnamefont {Yashar}\ \bibnamefont {Komijani}}, \bibinfo
  {author} {\bibfnamefont {Michael}\ \bibnamefont {Nicklas}}, \bibinfo {author}
  {\bibfnamefont {Robert}\ \bibnamefont {Borth}}, \bibinfo {author}
  {\bibfnamefont {An}~\bibnamefont {Wang}}, \bibinfo {author} {\bibfnamefont
  {Ye}~\bibnamefont {Chen}}, \bibinfo {author} {\bibfnamefont {Zhiyong}\
  \bibnamefont {Nie}}, \bibinfo {author} {\bibfnamefont {Rui}\ \bibnamefont
  {Li}}, \bibinfo {author} {\bibfnamefont {Xin}\ \bibnamefont {Lu}}, \bibinfo
  {author} {\bibfnamefont {Hanoh}\ \bibnamefont {Lee}}, \bibinfo {author}
  {\bibfnamefont {Michael}\ \bibnamefont {Smidman}}, \bibinfo {author}
  {\bibfnamefont {Frank}\ \bibnamefont {Steglich}}, \bibinfo {author}
  {\bibfnamefont {Piers}\ \bibnamefont {Coleman}}, \ and\ \bibinfo {author}
  {\bibfnamefont {Huiqiu}\ \bibnamefont {Yuan}},\ }\bibfield  {title} {\enquote
  {\bibinfo {title} {Strange-metal behaviour in a pure ferromagnetic kondo
  lattice},}\ }\href {\doibase 10.1038/s41586-020-2052-z} {\bibfield  {journal}
  {\bibinfo  {journal} {Nature}\ }\textbf {\bibinfo {volume} {579}},\ \bibinfo
  {pages} {51--55} (\bibinfo {year} {2020})}\BibitemShut {NoStop}%
\bibitem [{\citenamefont {Grissonnanche}\ \emph {et~al.}(2021)\citenamefont
  {Grissonnanche}, \citenamefont {Fang}, \citenamefont {Legros}, \citenamefont
  {Verret}, \citenamefont {Lalibert{\'e}}, \citenamefont {Collignon},
  \citenamefont {Zhou}, \citenamefont {Graf}, \citenamefont {Goddard},
  \citenamefont {Taillefer},\ and\ \citenamefont {Ramshaw}}]{2021natsm}%
  \BibitemOpen
  \bibfield  {author} {\bibinfo {author} {\bibfnamefont {Ga{\"e}l}\
  \bibnamefont {Grissonnanche}}, \bibinfo {author} {\bibfnamefont {Yawen}\
  \bibnamefont {Fang}}, \bibinfo {author} {\bibfnamefont {Ana{\"e}lle}\
  \bibnamefont {Legros}}, \bibinfo {author} {\bibfnamefont {Simon}\
  \bibnamefont {Verret}}, \bibinfo {author} {\bibfnamefont {Francis}\
  \bibnamefont {Lalibert{\'e}}}, \bibinfo {author} {\bibfnamefont
  {Cl{\'e}ment}\ \bibnamefont {Collignon}}, \bibinfo {author} {\bibfnamefont
  {Jianshi}\ \bibnamefont {Zhou}}, \bibinfo {author} {\bibfnamefont {David}\
  \bibnamefont {Graf}}, \bibinfo {author} {\bibfnamefont {Paul~A.}\
  \bibnamefont {Goddard}}, \bibinfo {author} {\bibfnamefont {Louis}\
  \bibnamefont {Taillefer}}, \ and\ \bibinfo {author} {\bibfnamefont {B.~J.}\
  \bibnamefont {Ramshaw}},\ }\bibfield  {title} {\enquote {\bibinfo {title}
  {Linear-in temperature resistivity from an isotropic planckian scattering
  rate},}\ }\href {\doibase 10.1038/s41586-021-03697-8} {\bibfield  {journal}
  {\bibinfo  {journal} {Nature}\ }\textbf {\bibinfo {volume} {595}},\ \bibinfo
  {pages} {667--672} (\bibinfo {year} {2021})}\BibitemShut {NoStop}%
\bibitem [{\citenamefont {Chen}\ \emph {et~al.}(2023)\citenamefont {Chen},
  \citenamefont {Lowder}, \citenamefont {Bakali}, \citenamefont {Andrews},
  \citenamefont {Schrenk}, \citenamefont {Waas}, \citenamefont {Svagera},
  \citenamefont {Eguchi}, \citenamefont {Prochaska}, \citenamefont {Wang},
  \citenamefont {Setty}, \citenamefont {Sur}, \citenamefont {Si}, \citenamefont
  {Paschen},\ and\ \citenamefont {Natelson}}]{2023scistrange}%
  \BibitemOpen
  \bibfield  {author} {\bibinfo {author} {\bibfnamefont {Liyang}\ \bibnamefont
  {Chen}}, \bibinfo {author} {\bibfnamefont {Dale~T.}\ \bibnamefont {Lowder}},
  \bibinfo {author} {\bibfnamefont {Emine}\ \bibnamefont {Bakali}}, \bibinfo
  {author} {\bibfnamefont {Aaron~Maxwell}\ \bibnamefont {Andrews}}, \bibinfo
  {author} {\bibfnamefont {Werner}\ \bibnamefont {Schrenk}}, \bibinfo {author}
  {\bibfnamefont {Monika}\ \bibnamefont {Waas}}, \bibinfo {author}
  {\bibfnamefont {Robert}\ \bibnamefont {Svagera}}, \bibinfo {author}
  {\bibfnamefont {Gaku}\ \bibnamefont {Eguchi}}, \bibinfo {author}
  {\bibfnamefont {Lukas}\ \bibnamefont {Prochaska}}, \bibinfo {author}
  {\bibfnamefont {Yiming}\ \bibnamefont {Wang}}, \bibinfo {author}
  {\bibfnamefont {Chandan}\ \bibnamefont {Setty}}, \bibinfo {author}
  {\bibfnamefont {Shouvik}\ \bibnamefont {Sur}}, \bibinfo {author}
  {\bibfnamefont {Qimiao}\ \bibnamefont {Si}}, \bibinfo {author} {\bibfnamefont
  {Silke}\ \bibnamefont {Paschen}}, \ and\ \bibinfo {author} {\bibfnamefont
  {Douglas}\ \bibnamefont {Natelson}},\ }\bibfield  {title} {\enquote {\bibinfo
  {title} {Shot noise in a strange metal},}\ }\href {\doibase
  10.1126/science.abq6100} {\bibfield  {journal} {\bibinfo  {journal}
  {Science}\ }\textbf {\bibinfo {volume} {382}},\ \bibinfo {pages} {907--911}
  (\bibinfo {year} {2023})},\ \Eprint
  {http://arxiv.org/abs/https://www.science.org/doi/pdf/10.1126/science.abq6100}
  {https://www.science.org/doi/pdf/10.1126/science.abq6100} \BibitemShut
  {NoStop}%
\bibitem [{\citenamefont {Li}\ and\ \citenamefont
  {Zhang}(2023)}]{2023npstrange}%
  \BibitemOpen
  \bibfield  {author} {\bibinfo {author} {\bibfnamefont {Lu}~\bibnamefont
  {Li}}\ and\ \bibinfo {author} {\bibfnamefont {Dechen}\ \bibnamefont
  {Zhang}},\ }\bibfield  {title} {\enquote {\bibinfo {title} {Probes to entropy
  flow in strange metals},}\ }\href {\doibase 10.1038/s41567-023-01981-0}
  {\bibfield  {journal} {\bibinfo  {journal} {Nature Physics}\ }\textbf
  {\bibinfo {volume} {19}},\ \bibinfo {pages} {307--308} (\bibinfo {year}
  {2023})}\BibitemShut {NoStop}%
\bibitem [{\citenamefont {Nguyen}\ \emph {et~al.}(2021)\citenamefont {Nguyen},
  \citenamefont {Sidorenko}, \citenamefont {Taupin}, \citenamefont {Knebel},
  \citenamefont {Lapertot}, \citenamefont {Schuberth},\ and\ \citenamefont
  {Paschen}}]{2021ncstrange}%
  \BibitemOpen
  \bibfield  {author} {\bibinfo {author} {\bibfnamefont {D.~H.}\ \bibnamefont
  {Nguyen}}, \bibinfo {author} {\bibfnamefont {A.}~\bibnamefont {Sidorenko}},
  \bibinfo {author} {\bibfnamefont {M.}~\bibnamefont {Taupin}}, \bibinfo
  {author} {\bibfnamefont {G.}~\bibnamefont {Knebel}}, \bibinfo {author}
  {\bibfnamefont {G.}~\bibnamefont {Lapertot}}, \bibinfo {author}
  {\bibfnamefont {E.}~\bibnamefont {Schuberth}}, \ and\ \bibinfo {author}
  {\bibfnamefont {S.}~\bibnamefont {Paschen}},\ }\bibfield  {title} {\enquote
  {\bibinfo {title} {Superconductivity in an extreme strange metal},}\ }\href
  {\doibase 10.1038/s41467-021-24670-z} {\bibfield  {journal} {\bibinfo
  {journal} {Nature Communications}\ }\textbf {\bibinfo {volume} {12}},\
  \bibinfo {pages} {4341} (\bibinfo {year} {2021})}\BibitemShut {NoStop}%
\bibitem [{\citenamefont {Yang}\ \emph {et~al.}(2023)\citenamefont {Yang},
  \citenamefont {Tao}, \citenamefont {Fang}, \citenamefont {Tang},
  \citenamefont {Yao}, \citenamefont {Yan}, \citenamefont {Jiang},
  \citenamefont {Xu}, \citenamefont {Huang}, \citenamefont {Ding},
  \citenamefont {Wang}, \citenamefont {Mao}, \citenamefont {Xing},\ and\
  \citenamefont {Xu}}]{2023npcuprate}%
  \BibitemOpen
  \bibfield  {author} {\bibinfo {author} {\bibfnamefont {Yusen}\ \bibnamefont
  {Yang}}, \bibinfo {author} {\bibfnamefont {Qian}\ \bibnamefont {Tao}},
  \bibinfo {author} {\bibfnamefont {Yuqiang}\ \bibnamefont {Fang}}, \bibinfo
  {author} {\bibfnamefont {Guoxiong}\ \bibnamefont {Tang}}, \bibinfo {author}
  {\bibfnamefont {Chao}\ \bibnamefont {Yao}}, \bibinfo {author} {\bibfnamefont
  {Xiaoxian}\ \bibnamefont {Yan}}, \bibinfo {author} {\bibfnamefont {Chenxi}\
  \bibnamefont {Jiang}}, \bibinfo {author} {\bibfnamefont {Xiangfan}\
  \bibnamefont {Xu}}, \bibinfo {author} {\bibfnamefont {Fuqiang}\ \bibnamefont
  {Huang}}, \bibinfo {author} {\bibfnamefont {Wenxin}\ \bibnamefont {Ding}},
  \bibinfo {author} {\bibfnamefont {Yu}~\bibnamefont {Wang}}, \bibinfo {author}
  {\bibfnamefont {Zhiqiang}\ \bibnamefont {Mao}}, \bibinfo {author}
  {\bibfnamefont {Hui}\ \bibnamefont {Xing}}, \ and\ \bibinfo {author}
  {\bibfnamefont {Zhu-An}\ \bibnamefont {Xu}},\ }\bibfield  {title} {\enquote
  {\bibinfo {title} {Anomalous enhancement of the nernst effect at the
  crossover between a fermi liquid and a strange metal},}\ }\href {\doibase
  10.1038/s41567-022-01904-5} {\bibfield  {journal} {\bibinfo  {journal}
  {Nature Physics}\ }\textbf {\bibinfo {volume} {19}},\ \bibinfo {pages}
  {379--385} (\bibinfo {year} {2023})}\BibitemShut {NoStop}%
\bibitem [{\citenamefont {Legros}\ \emph {et~al.}(2019)\citenamefont {Legros},
  \citenamefont {Benhabib}, \citenamefont {Tabis}, \citenamefont
  {Lalibert{\'e}}, \citenamefont {Dion}, \citenamefont {Lizaire}, \citenamefont
  {Vignolle}, \citenamefont {Vignolles}, \citenamefont {Raffy}, \citenamefont
  {Li}, \citenamefont {Auban-Senzier}, \citenamefont {Doiron-Leyraud},
  \citenamefont {Fournier}, \citenamefont {Colson}, \citenamefont {Taillefer},\
  and\ \citenamefont {Proust}}]{2019npcuprate}%
  \BibitemOpen
  \bibfield  {author} {\bibinfo {author} {\bibfnamefont {A.}~\bibnamefont
  {Legros}}, \bibinfo {author} {\bibfnamefont {S.}~\bibnamefont {Benhabib}},
  \bibinfo {author} {\bibfnamefont {W.}~\bibnamefont {Tabis}}, \bibinfo
  {author} {\bibfnamefont {F.}~\bibnamefont {Lalibert{\'e}}}, \bibinfo {author}
  {\bibfnamefont {M.}~\bibnamefont {Dion}}, \bibinfo {author} {\bibfnamefont
  {M.}~\bibnamefont {Lizaire}}, \bibinfo {author} {\bibfnamefont
  {B.}~\bibnamefont {Vignolle}}, \bibinfo {author} {\bibfnamefont
  {D.}~\bibnamefont {Vignolles}}, \bibinfo {author} {\bibfnamefont
  {H.}~\bibnamefont {Raffy}}, \bibinfo {author} {\bibfnamefont {Z.~Z.}\
  \bibnamefont {Li}}, \bibinfo {author} {\bibfnamefont {P.}~\bibnamefont
  {Auban-Senzier}}, \bibinfo {author} {\bibfnamefont {N.}~\bibnamefont
  {Doiron-Leyraud}}, \bibinfo {author} {\bibfnamefont {P.}~\bibnamefont
  {Fournier}}, \bibinfo {author} {\bibfnamefont {D.}~\bibnamefont {Colson}},
  \bibinfo {author} {\bibfnamefont {L.}~\bibnamefont {Taillefer}}, \ and\
  \bibinfo {author} {\bibfnamefont {C.}~\bibnamefont {Proust}},\ }\bibfield
  {title} {\enquote {\bibinfo {title} {Universal t-linear resistivity and
  planckian dissipation in overdoped cuprates},}\ }\href {\doibase
  10.1038/s41567-018-0334-2} {\bibfield  {journal} {\bibinfo  {journal} {Nature
  Physics}\ }\textbf {\bibinfo {volume} {15}},\ \bibinfo {pages} {142--147}
  (\bibinfo {year} {2019})}\BibitemShut {NoStop}%
\bibitem [{\citenamefont {Jiang}\ \emph {et~al.}(2023)\citenamefont {Jiang},
  \citenamefont {Qin}, \citenamefont {Wei}, \citenamefont {Xu}, \citenamefont
  {Ke}, \citenamefont {Zhu}, \citenamefont {Zhang}, \citenamefont {Zhao},
  \citenamefont {Liang}, \citenamefont {Wei}, \citenamefont {Lin},
  \citenamefont {Feng}, \citenamefont {Chen}, \citenamefont {Xiong},
  \citenamefont {Yuan}, \citenamefont {Zhu}, \citenamefont {Li}, \citenamefont
  {Xi}, \citenamefont {Wang}, \citenamefont {Yang}, \citenamefont {Wang},
  \citenamefont {Xiang}, \citenamefont {Hu}, \citenamefont {Jiang},
  \citenamefont {Chen}, \citenamefont {Jin},\ and\ \citenamefont
  {Zhao}}]{2023natiron}%
  \BibitemOpen
  \bibfield  {author} {\bibinfo {author} {\bibfnamefont {Xingyu}\ \bibnamefont
  {Jiang}}, \bibinfo {author} {\bibfnamefont {Mingyang}\ \bibnamefont {Qin}},
  \bibinfo {author} {\bibfnamefont {Xinjian}\ \bibnamefont {Wei}}, \bibinfo
  {author} {\bibfnamefont {Li}~\bibnamefont {Xu}}, \bibinfo {author}
  {\bibfnamefont {Jiezun}\ \bibnamefont {Ke}}, \bibinfo {author} {\bibfnamefont
  {Haipeng}\ \bibnamefont {Zhu}}, \bibinfo {author} {\bibfnamefont {Ruozhou}\
  \bibnamefont {Zhang}}, \bibinfo {author} {\bibfnamefont {Zhanyi}\
  \bibnamefont {Zhao}}, \bibinfo {author} {\bibfnamefont {Qimei}\ \bibnamefont
  {Liang}}, \bibinfo {author} {\bibfnamefont {Zhongxu}\ \bibnamefont {Wei}},
  \bibinfo {author} {\bibfnamefont {Zefeng}\ \bibnamefont {Lin}}, \bibinfo
  {author} {\bibfnamefont {Zhongpei}\ \bibnamefont {Feng}}, \bibinfo {author}
  {\bibfnamefont {Fucong}\ \bibnamefont {Chen}}, \bibinfo {author}
  {\bibfnamefont {Peiyu}\ \bibnamefont {Xiong}}, \bibinfo {author}
  {\bibfnamefont {Jie}\ \bibnamefont {Yuan}}, \bibinfo {author} {\bibfnamefont
  {Beiyi}\ \bibnamefont {Zhu}}, \bibinfo {author} {\bibfnamefont {Yangmu}\
  \bibnamefont {Li}}, \bibinfo {author} {\bibfnamefont {Chuanying}\
  \bibnamefont {Xi}}, \bibinfo {author} {\bibfnamefont {Zhaosheng}\
  \bibnamefont {Wang}}, \bibinfo {author} {\bibfnamefont {Ming}\ \bibnamefont
  {Yang}}, \bibinfo {author} {\bibfnamefont {Junfeng}\ \bibnamefont {Wang}},
  \bibinfo {author} {\bibfnamefont {Tao}\ \bibnamefont {Xiang}}, \bibinfo
  {author} {\bibfnamefont {Jiangping}\ \bibnamefont {Hu}}, \bibinfo {author}
  {\bibfnamefont {Kun}\ \bibnamefont {Jiang}}, \bibinfo {author} {\bibfnamefont
  {Qihong}\ \bibnamefont {Chen}}, \bibinfo {author} {\bibfnamefont {Kui}\
  \bibnamefont {Jin}}, \ and\ \bibinfo {author} {\bibfnamefont {Zhongxian}\
  \bibnamefont {Zhao}},\ }\bibfield  {title} {\enquote {\bibinfo {title}
  {Interplay between superconductivity and the strange-metal state in fese},}\
  }\href {\doibase 10.1038/s41567-022-01894-4} {\bibfield  {journal} {\bibinfo
  {journal} {Nature Physics}\ }\textbf {\bibinfo {volume} {19}},\ \bibinfo
  {pages} {365--371} (\bibinfo {year} {2023})}\BibitemShut {NoStop}%
\bibitem [{\citenamefont {Yang}\ \emph {et~al.}(2022)\citenamefont {Yang},
  \citenamefont {Liu}, \citenamefont {Liu}, \citenamefont {Wang}, \citenamefont
  {Qiu}, \citenamefont {Wang}, \citenamefont {Wang}, \citenamefont {He},
  \citenamefont {Li}, \citenamefont {Li}, \citenamefont {Tang}, \citenamefont
  {Wang}, \citenamefont {Xie}, \citenamefont {Valles}, \citenamefont {Xiong},\
  and\ \citenamefont {Li}}]{2022natBechgaard}%
  \BibitemOpen
  \bibfield  {author} {\bibinfo {author} {\bibfnamefont {Chao}\ \bibnamefont
  {Yang}}, \bibinfo {author} {\bibfnamefont {Haiwen}\ \bibnamefont {Liu}},
  \bibinfo {author} {\bibfnamefont {Yi}~\bibnamefont {Liu}}, \bibinfo {author}
  {\bibfnamefont {Jiandong}\ \bibnamefont {Wang}}, \bibinfo {author}
  {\bibfnamefont {Dong}\ \bibnamefont {Qiu}}, \bibinfo {author} {\bibfnamefont
  {Sishuang}\ \bibnamefont {Wang}}, \bibinfo {author} {\bibfnamefont {Yang}\
  \bibnamefont {Wang}}, \bibinfo {author} {\bibfnamefont {Qianmei}\
  \bibnamefont {He}}, \bibinfo {author} {\bibfnamefont {Xiuli}\ \bibnamefont
  {Li}}, \bibinfo {author} {\bibfnamefont {Peng}\ \bibnamefont {Li}}, \bibinfo
  {author} {\bibfnamefont {Yue}\ \bibnamefont {Tang}}, \bibinfo {author}
  {\bibfnamefont {Jian}\ \bibnamefont {Wang}}, \bibinfo {author} {\bibfnamefont
  {X.~C.}\ \bibnamefont {Xie}}, \bibinfo {author} {\bibfnamefont {James~M.}\
  \bibnamefont {Valles}}, \bibinfo {author} {\bibfnamefont {Jie}\ \bibnamefont
  {Xiong}}, \ and\ \bibinfo {author} {\bibfnamefont {Yanrong}\ \bibnamefont
  {Li}},\ }\bibfield  {title} {\enquote {\bibinfo {title} {Signatures of a
  strange metal in a bosonic system},}\ }\href {\doibase
  10.1038/s41586-021-04239-y} {\bibfield  {journal} {\bibinfo  {journal}
  {Nature}\ }\textbf {\bibinfo {volume} {601}},\ \bibinfo {pages} {205--210}
  (\bibinfo {year} {2022})}\BibitemShut {NoStop}%
\bibitem [{\citenamefont {Doiron-Leyraud}\ \emph {et~al.}(2009)\citenamefont
  {Doiron-Leyraud}, \citenamefont {Auban-Senzier}, \citenamefont {Ren\'e~de
  Cotret}, \citenamefont {Bourbonnais}, \citenamefont {J\'erome}, \citenamefont
  {Bechgaard},\ and\ \citenamefont {Taillefer}}]{2009prbBechgaard}%
  \BibitemOpen
  \bibfield  {author} {\bibinfo {author} {\bibfnamefont {Nicolas}\ \bibnamefont
  {Doiron-Leyraud}}, \bibinfo {author} {\bibfnamefont {Pascale}\ \bibnamefont
  {Auban-Senzier}}, \bibinfo {author} {\bibfnamefont {Samuel}\ \bibnamefont
  {Ren\'e~de Cotret}}, \bibinfo {author} {\bibfnamefont {Claude}\ \bibnamefont
  {Bourbonnais}}, \bibinfo {author} {\bibfnamefont {Denis}\ \bibnamefont
  {J\'erome}}, \bibinfo {author} {\bibfnamefont {Klaus}\ \bibnamefont
  {Bechgaard}}, \ and\ \bibinfo {author} {\bibfnamefont {Louis}\ \bibnamefont
  {Taillefer}},\ }\bibfield  {title} {\enquote {\bibinfo {title} {Correlation
  between linear resistivity and ${T}_{c}$ in the bechgaard salts and the
  pnictide superconductor
  $\text{Ba}{({\text{Fe}}_{1\ensuremath{-}x}{\text{Co}}_{x})}_{2}{\text{as}}_{2}$},}\
  }\href {\doibase 10.1103/PhysRevB.80.214531} {\bibfield  {journal} {\bibinfo
  {journal} {Phys. Rev. B}\ }\textbf {\bibinfo {volume} {80}},\ \bibinfo
  {pages} {214531} (\bibinfo {year} {2009})}\BibitemShut {NoStop}%
\bibitem [{\citenamefont {Aji}\ and\ \citenamefont
  {Varma}(2007)}]{prlphase_coherence}%
  \BibitemOpen
  \bibfield  {author} {\bibinfo {author} {\bibfnamefont {Vivek}\ \bibnamefont
  {Aji}}\ and\ \bibinfo {author} {\bibfnamefont {C.~M.}\ \bibnamefont
  {Varma}},\ }\bibfield  {title} {\enquote {\bibinfo {title} {Theory of the
  quantum critical fluctuations in cuprate superconductors},}\ }\href {\doibase
  10.1103/PhysRevLett.99.067003} {\bibfield  {journal} {\bibinfo  {journal}
  {Phys. Rev. Lett.}\ }\textbf {\bibinfo {volume} {99}},\ \bibinfo {pages}
  {067003} (\bibinfo {year} {2007})}\BibitemShut {NoStop}%
\bibitem [{\citenamefont {Zhu}\ \emph {et~al.}(2015)\citenamefont {Zhu},
  \citenamefont {Chen},\ and\ \citenamefont {Varma}}]{prbvortex_pairs}%
  \BibitemOpen
  \bibfield  {author} {\bibinfo {author} {\bibfnamefont {Lijun}\ \bibnamefont
  {Zhu}}, \bibinfo {author} {\bibfnamefont {Yan}\ \bibnamefont {Chen}}, \ and\
  \bibinfo {author} {\bibfnamefont {Chandra~M.}\ \bibnamefont {Varma}},\
  }\bibfield  {title} {\enquote {\bibinfo {title} {Local quantum criticality in
  the two-dimensional dissipative quantum xy model},}\ }\href {\doibase
  10.1103/PhysRevB.91.205129} {\bibfield  {journal} {\bibinfo  {journal} {Phys.
  Rev. B}\ }\textbf {\bibinfo {volume} {91}},\ \bibinfo {pages} {205129}
  (\bibinfo {year} {2015})}\BibitemShut {NoStop}%
\bibitem [{\citenamefont {Lee}\ \emph {et~al.}(2023{\natexlab{a}})\citenamefont
  {Lee}, \citenamefont {Kim},\ and\ \citenamefont {Kwon}}]{Lee1}%
  \BibitemOpen
  \bibfield  {author} {\bibinfo {author} {\bibfnamefont {Sukbae}\ \bibnamefont
  {Lee}}, \bibinfo {author} {\bibfnamefont {Ji-Hoon}\ \bibnamefont {Kim}}, \
  and\ \bibinfo {author} {\bibfnamefont {Young-Wan}\ \bibnamefont {Kwon}},\
  }\href@noop {} {\enquote {\bibinfo {title} {The first room-temperature
  ambient-pressure superconductor},}\ } (\bibinfo {year}
  {2023}{\natexlab{a}}),\ \Eprint {http://arxiv.org/abs/2307.12008}
  {arXiv:2307.12008 [cond-mat.supr-con]} \BibitemShut {NoStop}%
\bibitem [{\citenamefont {Lee}\ \emph {et~al.}(2023{\natexlab{b}})\citenamefont
  {Lee}, \citenamefont {Kim}, \citenamefont {Kim}, \citenamefont {Im},
  \citenamefont {An},\ and\ \citenamefont {Auh}}]{Lee2}%
  \BibitemOpen
  \bibfield  {author} {\bibinfo {author} {\bibfnamefont {Sukbae}\ \bibnamefont
  {Lee}}, \bibinfo {author} {\bibfnamefont {Jihoon}\ \bibnamefont {Kim}},
  \bibinfo {author} {\bibfnamefont {Hyun-Tak}\ \bibnamefont {Kim}}, \bibinfo
  {author} {\bibfnamefont {Sungyeon}\ \bibnamefont {Im}}, \bibinfo {author}
  {\bibfnamefont {SooMin}\ \bibnamefont {An}}, \ and\ \bibinfo {author}
  {\bibfnamefont {Keun~Ho}\ \bibnamefont {Auh}},\ }\href@noop {} {\enquote
  {\bibinfo {title} {Superconductor pb$_{10-x}$cu$_x$(po$_4$)$_6$o showing
  levitation at room temperature and atmospheric pressure and mechanism},}\ }
  (\bibinfo {year} {2023}{\natexlab{b}}),\ \Eprint
  {http://arxiv.org/abs/2307.12037} {arXiv:2307.12037 [cond-mat.supr-con]}
  \BibitemShut {NoStop}%
\bibitem [{\citenamefont {Wang}\ \emph {et~al.}(2023)\citenamefont {Wang},
  \citenamefont {Liu}, \citenamefont {Ge}, \citenamefont {Ji}, \citenamefont
  {Ji}, \citenamefont {Liu}, \citenamefont {Ai}, \citenamefont {Ma},
  \citenamefont {Qi},\ and\ \citenamefont {Wang}}]{Wang2023}%
  \BibitemOpen
  \bibfield  {author} {\bibinfo {author} {\bibfnamefont {Pinyuan}\ \bibnamefont
  {Wang}}, \bibinfo {author} {\bibfnamefont {Xiaoqi}\ \bibnamefont {Liu}},
  \bibinfo {author} {\bibfnamefont {Jun}\ \bibnamefont {Ge}}, \bibinfo {author}
  {\bibfnamefont {Chengcheng}\ \bibnamefont {Ji}}, \bibinfo {author}
  {\bibfnamefont {Haoran}\ \bibnamefont {Ji}}, \bibinfo {author} {\bibfnamefont
  {Yanzhao}\ \bibnamefont {Liu}}, \bibinfo {author} {\bibfnamefont {Yiwen}\
  \bibnamefont {Ai}}, \bibinfo {author} {\bibfnamefont {Gaoxing}\ \bibnamefont
  {Ma}}, \bibinfo {author} {\bibfnamefont {Shichao}\ \bibnamefont {Qi}}, \ and\
  \bibinfo {author} {\bibfnamefont {Jian}\ \bibnamefont {Wang}},\ }\bibfield
  {title} {\enquote {\bibinfo {title} {Ferromagnetic and insulating behavior in
  both half magnetic levitation and non-levitation lk-99 like samples},}\
  }\href {\doibase 10.1007/s44214-023-00035-z} {\bibfield  {journal} {\bibinfo
  {journal} {Quantum Frontiers}\ }\textbf {\bibinfo {volume} {2}},\ \bibinfo
  {pages} {10} (\bibinfo {year} {2023})}\BibitemShut {NoStop}%
\bibitem [{\citenamefont {Guo}\ \emph {et~al.}(2023)\citenamefont {Guo},
  \citenamefont {Li},\ and\ \citenamefont {Jia}}]{Guo2023}%
  \BibitemOpen
  \bibfield  {author} {\bibinfo {author} {\bibfnamefont {Kaizhen}\ \bibnamefont
  {Guo}}, \bibinfo {author} {\bibfnamefont {Yuan}\ \bibnamefont {Li}}, \ and\
  \bibinfo {author} {\bibfnamefont {Shuang}\ \bibnamefont {Jia}},\ }\bibfield
  {title} {\enquote {\bibinfo {title} {Ferromagnetic half levitation of
  lk-99-like synthetic samples},}\ }\href {\doibase 10.1007/s11433-023-2201-9}
  {\bibfield  {journal} {\bibinfo  {journal} {Science China Physics, Mechanics
  \& Astronomy}\ }\textbf {\bibinfo {volume} {66}},\ \bibinfo {pages} {107411}
  (\bibinfo {year} {2023})}\BibitemShut {NoStop}%
\bibitem [{\citenamefont {Habamahoro}\ \emph {et~al.}(2023)\citenamefont
  {Habamahoro}, \citenamefont {Bontke}, \citenamefont {Chirom}, \citenamefont
  {Wu}, \citenamefont {Bao}, \citenamefont {Deng},\ and\ \citenamefont
  {Chu}}]{2023habamahoro}%
  \BibitemOpen
  \bibfield  {author} {\bibinfo {author} {\bibfnamefont {T.}~\bibnamefont
  {Habamahoro}}, \bibinfo {author} {\bibfnamefont {T.}~\bibnamefont {Bontke}},
  \bibinfo {author} {\bibfnamefont {M.}~\bibnamefont {Chirom}}, \bibinfo
  {author} {\bibfnamefont {Z.}~\bibnamefont {Wu}}, \bibinfo {author}
  {\bibfnamefont {J.~M.}\ \bibnamefont {Bao}}, \bibinfo {author} {\bibfnamefont
  {L.~Z.}\ \bibnamefont {Deng}}, \ and\ \bibinfo {author} {\bibfnamefont
  {C.~W.}\ \bibnamefont {Chu}},\ }\href@noop {} {\enquote {\bibinfo {title}
  {Replication and study of anomalies in lk-99--the alleged ambient-pressure,
  room-temperature superconductor},}\ } (\bibinfo {year} {2023}),\ \Eprint
  {http://arxiv.org/abs/2311.03558} {arXiv:2311.03558 [cond-mat.supr-con]}
  \BibitemShut {NoStop}%
\bibitem [{\citenamefont {Liu}\ \emph {et~al.}(2023{\natexlab{a}})\citenamefont
  {Liu}, \citenamefont {He}, \citenamefont {Peng}, \citenamefont {Zhen},
  \citenamefont {Chen}, \citenamefont {Wang}, \citenamefont {Yang},
  \citenamefont {Qiao}, \citenamefont {Yao},\ and\ \citenamefont
  {Ma}}]{liu2023longcoherence}%
  \BibitemOpen
  \bibfield  {author} {\bibinfo {author} {\bibfnamefont {Jicheng}\ \bibnamefont
  {Liu}}, \bibinfo {author} {\bibfnamefont {Chenao}\ \bibnamefont {He}},
  \bibinfo {author} {\bibfnamefont {Yin-Hui}\ \bibnamefont {Peng}}, \bibinfo
  {author} {\bibfnamefont {Zhihao}\ \bibnamefont {Zhen}}, \bibinfo {author}
  {\bibfnamefont {Guanhua}\ \bibnamefont {Chen}}, \bibinfo {author}
  {\bibfnamefont {Jia}\ \bibnamefont {Wang}}, \bibinfo {author} {\bibfnamefont
  {Xiao-Bao}\ \bibnamefont {Yang}}, \bibinfo {author} {\bibfnamefont
  {Xianfeng}\ \bibnamefont {Qiao}}, \bibinfo {author} {\bibfnamefont {Yao}\
  \bibnamefont {Yao}}, \ and\ \bibinfo {author} {\bibfnamefont {Dongge}\
  \bibnamefont {Ma}},\ }\href@noop {} {\enquote {\bibinfo {title}
  {Long-coherence pairing of low-mass conduction electrons in
  copper-substituted lead apatite},}\ } (\bibinfo {year}
  {2023}{\natexlab{a}}),\ \Eprint {http://arxiv.org/abs/2310.17160}
  {arXiv:2310.17160 [quant-ph]} \BibitemShut {NoStop}%
\bibitem [{\citenamefont {Liu}\ \emph {et~al.}(2023{\natexlab{b}})\citenamefont
  {Liu}, \citenamefont {He}, \citenamefont {Huang}, \citenamefont {Zhen},
  \citenamefont {Chen}, \citenamefont {Luo}, \citenamefont {Qiao},
  \citenamefont {Yao},\ and\ \citenamefont {Ma}}]{2023lowfield}%
  \BibitemOpen
  \bibfield  {author} {\bibinfo {author} {\bibfnamefont {Jicheng}\ \bibnamefont
  {Liu}}, \bibinfo {author} {\bibfnamefont {Chenao}\ \bibnamefont {He}},
  \bibinfo {author} {\bibfnamefont {Weijie}\ \bibnamefont {Huang}}, \bibinfo
  {author} {\bibfnamefont {Zhihao}\ \bibnamefont {Zhen}}, \bibinfo {author}
  {\bibfnamefont {Guanhua}\ \bibnamefont {Chen}}, \bibinfo {author}
  {\bibfnamefont {Tianyong}\ \bibnamefont {Luo}}, \bibinfo {author}
  {\bibfnamefont {Xianfeng}\ \bibnamefont {Qiao}}, \bibinfo {author}
  {\bibfnamefont {Yao}\ \bibnamefont {Yao}}, \ and\ \bibinfo {author}
  {\bibfnamefont {Dongge}\ \bibnamefont {Ma}},\ }\href@noop {} {\enquote
  {\bibinfo {title} {Strange memory effect of low-field microwave absorption in
  copper-substituted lead apatite},}\ } (\bibinfo {year}
  {2023}{\natexlab{b}}),\ \Eprint {http://arxiv.org/abs/2312.10391}
  {arXiv:2312.10391 [cond-mat.supr-con]} \BibitemShut {NoStop}%
\bibitem [{\citenamefont {Wang}\ \emph {et~al.}(2024)\citenamefont {Wang},
  \citenamefont {Yao}, \citenamefont {Shi}, \citenamefont {Zhao}, \citenamefont
  {Wu}, \citenamefont {Wu}, \citenamefont {Geng}, \citenamefont {Ye},\ and\
  \citenamefont {Chen}}]{wang2024possible}%
  \BibitemOpen
  \bibfield  {author} {\bibinfo {author} {\bibfnamefont {Hongyang}\
  \bibnamefont {Wang}}, \bibinfo {author} {\bibfnamefont {Yao}\ \bibnamefont
  {Yao}}, \bibinfo {author} {\bibfnamefont {Ke}~\bibnamefont {Shi}}, \bibinfo
  {author} {\bibfnamefont {Yijing}\ \bibnamefont {Zhao}}, \bibinfo {author}
  {\bibfnamefont {Hao}\ \bibnamefont {Wu}}, \bibinfo {author} {\bibfnamefont
  {Zhixing}\ \bibnamefont {Wu}}, \bibinfo {author} {\bibfnamefont {Zhihui}\
  \bibnamefont {Geng}}, \bibinfo {author} {\bibfnamefont {Shufeng}\
  \bibnamefont {Ye}}, \ and\ \bibinfo {author} {\bibfnamefont {Ning}\
  \bibnamefont {Chen}},\ }\href@noop {} {\enquote {\bibinfo {title} {Possible
  meissner effect near room temperature in copper-substituted lead apatite},}\
  } (\bibinfo {year} {2024}),\ \Eprint {http://arxiv.org/abs/2401.00999}
  {arXiv:2401.00999 [cond-mat.supr-con]} \BibitemShut {NoStop}%
\bibitem [{\citenamefont {Bi}\ \emph {et~al.}(2022)\citenamefont {Bi},
  \citenamefont {Huang}, \citenamefont {Qin}, \citenamefont {Qiu},\ and\
  \citenamefont {Yuan}}]{2022cap}%
  \BibitemOpen
  \bibfield  {author} {\bibinfo {author} {\bibfnamefont {Xiang-Yu}\
  \bibnamefont {Bi}}, \bibinfo {author} {\bibfnamefont {Jun-Wei}\ \bibnamefont
  {Huang}}, \bibinfo {author} {\bibfnamefont {Feng}\ \bibnamefont {Qin}},
  \bibinfo {author} {\bibfnamefont {Cai-Yu}\ \bibnamefont {Qiu}}, \ and\
  \bibinfo {author} {\bibfnamefont {Hong-Tao}\ \bibnamefont {Yuan}},\
  }\bibfield  {title} {\enquote {\bibinfo {title} {Quantum oscillation
  phenomena in low-dimensional superconductors},}\ }\href {\doibase
  10.7498/aps.71.20212289} {\bibfield  {journal} {\bibinfo  {journal} {Acta
  Physica Sinica}\ }\textbf {\bibinfo {volume} {71}},\ \bibinfo {pages}
  {127402--1--127402--19} (\bibinfo {year} {2022})}\BibitemShut {NoStop}%
\bibitem [{\citenamefont {Deppe}\ \emph {et~al.}(2004)\citenamefont {Deppe},
  \citenamefont {Saito}, \citenamefont {Tanaka},\ and\ \citenamefont
  {Takayanagi}}]{2004cap}%
  \BibitemOpen
  \bibfield  {author} {\bibinfo {author} {\bibfnamefont {Frank}\ \bibnamefont
  {Deppe}}, \bibinfo {author} {\bibfnamefont {Shiro}\ \bibnamefont {Saito}},
  \bibinfo {author} {\bibfnamefont {Hirotaka}\ \bibnamefont {Tanaka}}, \ and\
  \bibinfo {author} {\bibfnamefont {Hideaki}\ \bibnamefont {Takayanagi}},\
  }\bibfield  {title} {\enquote {\bibinfo {title} {{Determination of the
  capacitance of nm scale Josephson junctions}},}\ }\href {\doibase
  10.1063/1.1645673} {\bibfield  {journal} {\bibinfo  {journal} {Journal of
  Applied Physics}\ }\textbf {\bibinfo {volume} {95}},\ \bibinfo {pages}
  {2607--2613} (\bibinfo {year} {2004})},\ \Eprint
  {http://arxiv.org/abs/https://pubs.aip.org/aip/jap/article-pdf/95/5/2607/18708172/2607\_1\_online.pdf}
  {https://pubs.aip.org/aip/jap/article-pdf/95/5/2607/18708172/2607\_1\_online.pdf}
  \BibitemShut {NoStop}%
\end{thebibliography}%
%\bibliography{ref}

\end{document}